\documentclass[twocolumn,showpacs,%
aps,superscriptaddress,%eqsecnum,
prd,notitlepage,showkeys,
nofootinbib]{revtex4-1}

\usepackage{ulem}
\usepackage{amssymb}
\usepackage{amsmath}
\usepackage{graphicx}
\usepackage{dcolumn}
\usepackage[colorlinks,urlcolor=blue,citecolor=blue,linkcolor=blue]{hyperref}
\usepackage{color,units}
\usepackage[dvipsnames]{xcolor} % for more colours!
\usepackage{lineno}
\usepackage{xspace}
\usepackage{longtable} 
\usepackage{float} 
\usepackage{multirow}
\usepackage{amsfonts,wasysym,epsfig,verbatim,subfigure,bm,mathrsfs,lipsum}
\begin{document}
\newcommand{\IUCAA}{Inter-University Centre for Astronomy and
Astrophysics, Post Bag 4, Ganeshkhind, Pune 411 007, India}

\newcommand{\MPI}{Max-Planck-Institut f{\"u}r Gravitationsphysik (Albert-Einstein-Institut), D-30167 Hannover, Germany}

\newcommand{\LBNZ}{Leibniz Universit{\"a}t Hannover, D-30167 Hannover, Germany}

\title{Eccentricity-tide coupling: impact on binary neutron stars and extreme mass-ratio inspirals}
\author{J. P. Bernaldez}\email{john.bernaldez@aei.mpg.de}

\author{Sayak Datta}\email{sayak.datta@aei.mpg.de}
\affiliation{\MPI}\affiliation{\LBNZ}
\date{\today}
%\author{Bernaldez, J.P. and Datta, S.}
\date{\today}
\newcommand{\sayak}[1]{{\color{red}\bf  SD:} {\color{red} #1}}
\newcommand{\jp}[1]{{\color{orange} #1}}
%\begin{document}

\begin{abstract}
We study the effect of tidal interaction between two compact bodies in an eccentric orbit. We assume the tidal fields to be static. Therefore, we ignore the dynamic tides and resonant excitations. Using the results, we find the analytical expression for the phase shift of the emitted gravitational wave. In the process, we find that in the leading order, the initial eccentricity $e_0$ and the dimensionless tidal deformability $\Lambda$ couple as $\sim e_0^n\Lambda$, where $n$ is a positive number. We only focus on the dominant contribution, i.e., $e_0^2\Lambda$. We also compute the accumulated dephasing for binary neutron star systems. We find that for optimistic values of eccentricities $e_0 \sim .05$ and $\Lambda \sim 600$, the accumulated dephasing is $\mathcal{O}(10^{-4})$ radian, requiring a signal-to-noise ratio $\sim 7000$ to be observable. Therefore, these effects can be measured in binary neutron star systems with large eccentricities if the signal-to-noise ratios of the systems are also very large. Hence, in third-generation detectors, it may have an observable impact if the systems have large eccentricities. We also explore the impact of this effect on extreme mass-ratio inspirals (EMRIs). We find that even for supermassive bodies with small values of $\Lambda \sim 10^{-3}$, this effect has large dephasing in EMRIs $\sim \mathcal{O}(10)$ radian. Therefore, this effect will help in probing the nature of the supermassive bodies in an EMRI.
\end{abstract}

\maketitle

\section{Introduction}

The recent detection of gravitational waves (GWs) by LIGO~\cite{advanced-ligo}/Virgo~\cite{advanced-virgo} (LVC) collaboration from two binary neutron star (BNS) merger event named GW170817~\cite{TheLIGOScientific:2017qsa,Abbott:2018exr} and GW190425~\cite{Abbott:2020uma} has ushered in a new light for constraining the equation of state (EOS)
of neutron stars (NSs) \cite{Abbott:2018exr, LIGOScientific:2018hze, LIGOScientific:2017zic}. The tidally deformed components of a BNS merger in their late inspiral phase leave a detectable imprint on the emitted GW signals. These imprints depend on the EOS of neutron stars~\cite{Flanagan:2007ix, Hinderer:2007mb, Damour:2009vw, Binnington:2009bb}. The measurement of the tidal deformability associated with these observations, therefore, provides direct information about the equation of state of the NS~\cite{Flanagan:2007ix, Hinderer:2007mb, Damour:2009vw, Binnington:2009bb}. Thus, one can use this imprint to study the properties
of dense matter far from the nuclear saturation density. The tidal deformability of an NS does not only depend on the EOS of the matter inside the star but also on the fluid nature of the matter, and the symmetry of space-time~\cite{Char:2018grw, Datta:2019ueq, Biswas:2019gkw, Biswas:2019ifs}. Therefore, the information contained in the tidal deformability is very rich. Apart from GWs, other types of astrophysical observations also have supplied complementary constraints on the properties of NSs. Space-based mission led by Neutron Star Interior Composition ExploreR (NICER) collaboration has already presented a quite precise measurement of mass and radius of PSR J0030+0451~\cite{Miller:2019cac, Riley:2019yda} and PSR J0740+6620~\cite{Miller:2021qha, Riley:2021pdl} using pulse-profile modeling. Additionally, observations of massive pulsars~\cite{Antoniadis:2013pzd, Cromartie:2019kug,2021arXiv210400880F} by radio telescopes provide important insights on the high-density part of the NS EOS. All of these measurements of NS macroscopic properties, such as mass ($M$), radius ($R$), and tidal deformability ($\lambda$), lead us to the EOS which is the same for all the NSs present in the universe.

To measure these macroscopic properties, the waveform models have been constructed under the approximation of circular orbits. Theoretically, it is expected that BNSs get circularized by the time it enters the observable band. Even if it retains some eccentricity, it is expected to be small. The measurement of the eccentricities in the observed systems is consistent with such expectations \cite{Romero-Shaw:2020aaj, Lenon:2020oza}. For these reasons tidal interactions in BNSs got attention primarily for the circular orbit. In Ref. \cite{Flanagan:2007ix}, for the first time, the impact of the tidal deformability on the emitted GW from a circular orbit binary was calculated. Then, these results were extended, taking considerations of higher post-Newtonian corrections \cite{Bini:2012gu, Damour:2012yf, Maselli:2012zq, Vines:2011ud, Vines:2010ca}. The initial works primarily focused on the impact of the static tides. However, these were also extended to address the dynamic tides and resonant excitations \cite{Flanagan:2007ix, Steinhoff:2016rfi, Steinhoff:2021dsn, Schmidt:2019wrl, Pratten:2019sed}. In Ref. \cite{Pani:2015nua, Landry:2017piv, Abdelsalhin:2018reg, Castro:2022mpw}, the impact of the spin of NSs has been explored. The proposed 3rd generation (3G) GW detectors, Einstein Telescope (ET) and Cosmic Explorer (CE) will have
a higher sensitivity than current detectors, which will result in higher signal-to-noise ratio (SNR) for BNSs \cite{Reitze:2019iox, maggiore2020science}. In light of that, these effects are being explored for observational purposes. In Ref. \cite{Xu:2017hqo, Yang:2018bzx, Yang:2019kmf, Wang:2020iqj}, the tidal interactions in the presence of eccentricities have also been explored in a very general setting. These works primarily focused on numerically studying the properties of this interaction and computing resonant excitations.

In this paper, we focus on studying the impact of tidal deformability in eccentric orbits. However, we invest in a much simpler project. We would like to compute the impact of eccentricity in the presence of a static tide. Under such assumptions, we will derive analytical expressions for orbital energy, GW flux, and also the modified expression for phase. As a result, this expression can be used for parameter estimation purposes. Since it is expected that the orbits will get circularized when it enters the observable band, we expect that the eccentricity will contribute primarily in the earlier part of the inspiral. Therefore, the assumption of the static tide is very close to reality, as far as the BNSs are concerned. However, once the expressions are found, we will also explore the impact on other sources for exploration purposes. We work in the unit $G=c=1$.

\section{Tidal Love number}

Consider a static, spherically symmetric star of mass $m$ placed in an external quadrupolar tidal field $\mathcal{E}_{ij}$. In response to the external tidal field, the star will develop  an induced quadrupole moment ${Q}_{ij}$. The induced quadrupole moment depends on both the external tidal field and the properties of matter inside the star. This, as a result, affects the metric of the star. In the star’s local asymptotic rest frame (asymptotically mass centered Cartesian coordinates) at large $r$, the metric coefficient $g_{tt}$ is given by (in units with $G = c = 1$) :

\begin{equation}
    \frac{1-g_{tt}}{2} = -\frac{m}{r} -\frac{3Q_{ij}}{2r^3}(n^in^j -\frac{1}{3}\delta_{ij}) + \frac{1}{2}\mathcal{E}_{ij}x^ix^j,
\end{equation}
where $n^i = x^i/r$. This expansion is used to define the traceless tensors $Q_{ij}$ and the tidal field $\mathcal{E}_{ij}$. To linear order in static case, the induced quadrupole will be of the form

\begin{equation}\label{eq:TD definition}
    Q_{ij} = -\lambda \mathcal{E}_{ij}
\end{equation}

Here, $\lambda$ is a constant, namely tidal deformability. It is connected to the dimensionless tidal Love number $k_2 =\frac{3}{2} \lambda R^{-5}$, where $R$ is the radius of the star. In the linear approximation, the relation between $\mathcal{E}_{ij}$ and $Q_{ij}$ defines the tidal deformability $\lambda$. Hence, the details of the quadrupolar structure mimic the external tidal field, via $\mathcal{E}_{ij}$, and $\lambda$ captures the response of the body. This approximation, however, is valid only when the tide is considered to be static, i.e. orbital time scale is much larger than the time scale of the star's internal dynamics. This approximation breaks down in the late inspiral and necessitates including the dynamic nature of the tidal field. This is called the dynamic tide, which we will leave for future study. As the star's internal response is captured through $\lambda$, it heavily depends on the EOS of the star. The measurement of $\lambda$ from GW astronomy has already provided us with information regarding the EOS of the NSs. In the future, with more sensitive detectors, it is expected to provide us with further details in a precise manner. Therefore, it is of utmost importance to accurately model the contribution. In the next sections, we will explore these effects in a general setting.

\section{Equation of motion in the presence of a tidal field}

In this section, we will focus on deriving orbital equations of motions and their solutions. We consider a binary with component masses $m_i$ and tidal deformabilitties $\lambda_i$, where $i=1,2$. Similar to Ref. \cite{Flanagan:2007ix}, for simplicity, we compute only the effects on the first body. For the total contribution of both the bodies, a similar contribution can be added for the second body by interchanging the labels $(1\leftrightarrow 2)$. In the phase of the emitted GW, the signals from the two stars simply get added. The action of such systems can be expressed as,

\begin{equation}
    \begin{split}
        S = &\int dt \bigg[\frac{1}{2}\mu \dot r(t)^2+\frac{1}{2} \mu r(t)^2 \dot \phi(t)^2+\frac{M \mu}{r(t)}\bigg] - \Bigg\{\frac{1}{2}\int dt \, Q_{ij} \,\mathcal{E}_{ij}\\ &- \sum_n \int dt \frac{1}{4\lambda_{1,n} \omega_n^2} \bigg[\dot{Q}^{(n)}_{ij} \dot{Q}^{(n)}_{ij} - \omega_n^2Q^{(n)}_{ij} Q^{(n)}_{ij}\bigg] + 1\leftrightarrow2 \,\Bigg\},
    \end{split}
\end{equation}
where $M$ and $\mu$ are the total and reduced mass of the system, and $\mathcal{E}_{ij}=-m_2 \partial_i \partial_j (1/r)$ is the tidal field. $Q^{(n)}_{ij}$ and $\lambda_{1,n}$ are the contribution to the total induced quadrupole moment $Q_{ij}$ and tidal deformability, respectively. Alongside, $\lambda_1 \equiv \sum_n \lambda_{1,n}$ and $Q_{ij} \equiv \sum_n Q^{(n)}_{ij}$. The relative displacement is written as, ${\bf x} =(r\cos\phi,\, r\sin\phi, 0)$.

This form of action was considered in Ref. \cite{Flanagan:2007ix} to find the contributions of the different frequencies of the modes. However, the mode frequencies contribute in the very late inspiral stages. It is well known that the eccentricities of a binary system decreases with increasing frequency. Therefore, the dominant contribution in the eccentric orbit will come from the initial phase of the inspiral. The frequency $\omega$ in that phase is much smaller than the mode frequencies $\omega_n$ of the stars, i.e. $\omega_n \gg \omega$. For this purpose, we will ignore such contributions.

 Unlike in Ref. \cite{Flanagan:2007ix} we will vary the action w.r.t. $r, \, \phi,$ and $Q^{(n)}_{ij}$. The resulting equations are as follows,

\begin{eqnarray}
    \label{eq:orbit1}\mu \Ddot{r}(t) =\mu r(t) \, \dot\phi(t)^2-\frac{M \mu}{r(t)^2}-\frac{1}{2}Q_{ij}\frac{\partial \mathcal{E}_{ij}}{\partial r}\\
    \label{eq:phi}\frac{d}{dt}\big(\mu r^2 \dot{\phi}\big)=-\frac{1}{2}Q_{ij}\frac{\partial \mathcal{E}_{ij}}{\partial \phi} \\
    \label{eq:QM}\Ddot{Q}^{(n)}_{ij} + \omega_n^2 Q^{(n)}_{ij} = m_2\lambda_{1,n}~\omega_n^2\partial_i\partial_j\frac{1}{r},
\end{eqnarray}

In the current scenario, we consider that the mode frequency does not contribute. This eventually implies that the time variation of quadrupole moment is very small, i.e. $\Ddot{Q}^{(n)}_{ij} \ll \omega_n^2 Q^{(n)}_{ij}$. Therefore, Eq. (\ref{eq:QM}) simplifies to,

\begin{equation}
    Q^{(n)}_{ij} = m_2\lambda_{1,n}~\partial_i\partial_j\frac{1}{r}
\end{equation}
and, as a result, it boils down to Eq. (\ref{eq:TD definition}), $Q_{ij} = -\lambda \mathcal{E}_{ij}$. We will use this result through out the paper. With the solution of $ Q^{(n)}_{ij}$ Eq. (\ref{eq:orbit1}) can be simplified as follows:

\begin{equation}
    \label{eq:orbit}\mu \Ddot{r}(t) =\mu r(t) \, \dot\phi(t)^2-\frac{M \mu}{r(t)^2}-\frac{9 m_2^2 \lambda_{1}}{r(t)^7}
\end{equation}

In Eq. (\ref{eq:phi}), it is noteworthy that angular momentum is not always conserved. Its conservation lies on the structure of the induced quadrupole moment. However, in case of static tide due to Eq. (\ref{eq:TD definition}), the right-hand side of the equation vanishes and the angular momentum is conserved. However, this is not true in general and in the range $\omega \sim \omega_n$. This aspect has been explored in Ref. \cite{Yang:2019kmf}. 

It is easier to solve Eq. (\ref{eq:orbit}) by re-expressing the time variation of radius in terms of the time variation with $\phi$. Hence, we use,
\begin{equation}\label{22}
    \dot{r}=\frac{dr}{d \phi} \dot{\phi}.
\end{equation}

Using the above equation, Eq. (\ref{eq:orbit}) takes the following form,

\begin{equation}
    \label{eq:r}\frac{d^2r}{d \phi^2}\mu^2 r^4 \dot \phi^2   -2\mu^2 r^3 \dot \phi^2 \left(\frac{dr}{d \phi}\right)^2 =  \mu^2 r^5 \dot \phi^2-M \mu^2 r^2-\frac{9 \mu m_2^2 \lambda_{1}}{r^3}
\end{equation}

Except the last term in the right-hand side, the equation is similar to the equation of a gravitating body. The last term represents the leading order tidal interaction.

In the absence of a tidal field we can take the limit $(\lambda_{1} \rightarrow 0)$, and the solution satisfies an equation of ellipse,

\begin{equation}
    r = \frac{a(1-e^2)}{(1+e\cos\phi)},
\end{equation}
where $a$ and $e$ are the semi-major axis and eccentricity, respectively. We will assume that the effect of the tidal field is perturbative in nature. Therefore, the corresponding solution can be treated as a perturbation compared to the leading contribution. With this, we can replace $\dot{\phi}$ using the leading order expression of eccentric orbit, $\dot \phi=  \omega(1-e^2)^{-3/2}\big(1+e \cos(\phi)\big)^2$ \cite{PhysRev.131.435, Moore:2016qxz}.

Thus, we obtain:
\begin{widetext}
\begin{equation}\label{29}
    -(1-e^2)^3(9m_2^2 \lambda_{1}+M \mu r^5)+(1+e \cos \phi)^4\left(\mu \omega^2 r^8 +2 \mu \omega^2 r^6 \left(\dfrac{dr}{d\phi}\right)^2-\mu \omega^2 r^7 \dfrac{d^2r}{d\phi^2}\right)=0,
\end{equation}
\end{widetext}

where  $\omega = a^{-3/2} M^{1/2}$ is the average orbital frequency.

To separate out the perturbation, we use the ansatz:

\begin{equation}\label{30}
    r(\omega,\phi,e) = \frac{M^{1/3} \omega^{-2/3}(1-e^2)}{(1+e \cos{\phi})}(1+\gamma \, \bar{\delta r}(\phi,e)),
\end{equation}
where $\bar{\delta r}$ corresponds to perturbation in radius and $\gamma = \frac{3m_2^2 \lambda_{1}  \omega^{10/3}}{\mu M^{8/3} } $. Using the ansatz, we find the equation for $\bar{\delta r}$ to be:

\begin{widetext}
    \begin{equation}\label{31}
    (1-e^2)^5 (1+e \cos \phi) \frac{d^2\bar{\delta r}}{d\phi^2}-2 e(1-e^2)^5 \sin \phi \frac{d\bar{\delta r}}{d\phi}-3(1-e^2)^5\bar{\delta r}+3(1+e \cos \phi)^5=0
\end{equation}
\end{widetext}

We expand this equation in series of $e$ and keep up to the $e^2$ term. Then, it is straight forward to find the solution to be,

\begin{equation}\label{34}
    \bar{\delta r} = 1 + e \frac{15}{4} \cos{\phi} +e^2\left( \frac{85}{8}+\frac{75}{56}\cos{(2\phi)}\right).
\end{equation}

The final expression of $r$ with $e$ is given by
\begin{equation}\label{35}
    \begin{split}
        r(\omega,\phi,e) = &\frac{ M^{1/3} \omega^{-2/3}(1-e^2)}{(1+e \cos{\phi})}\Big\{1 +\gamma  \Big[1 + e \frac{15}{4} \cos{\phi}\\
        &+e^2\left( \frac{85}{8}+\frac{75}{56}\cos{2\phi}\right)\Big]\Big\}.
    \end{split}
\end{equation}
 This result will be used in the paper to find the GW contributions. Note, in the circular limit $(e=0)$, the result exactly reproduces the result derived in Ref. \cite{Flanagan:2007ix}.

\section{Impact on the emitted GW}

In the last section, we derived the solution of orbital motion in the presence of a tidal field. In this section, we want to compute the impact of this on the emitted GW from the system. The impact on the GW is computed by considering the change in the inspiral rate, which depends on the emitted GW flux from an orbit. With the loss of energy, the orbits shrink to reach a merger. Due to the tidal interaction, as the orbital structures get modified, so do their energy content and quadrupolar structure. The total energy of the system at an instant can be determined by computing the Hamiltonian $(H)$ of the system. Once, the Hamiltonian is known we compute an orbital average to find the average energy $E = \langle H \rangle.$

\begin{figure*}[ht]
\centering     %%% not \center

\subfigure[~$m_1=m_2=1.4 M_\odot$ and $q=1$ for $\Lambda=400$]{\label{}\includegraphics[width=85mm]{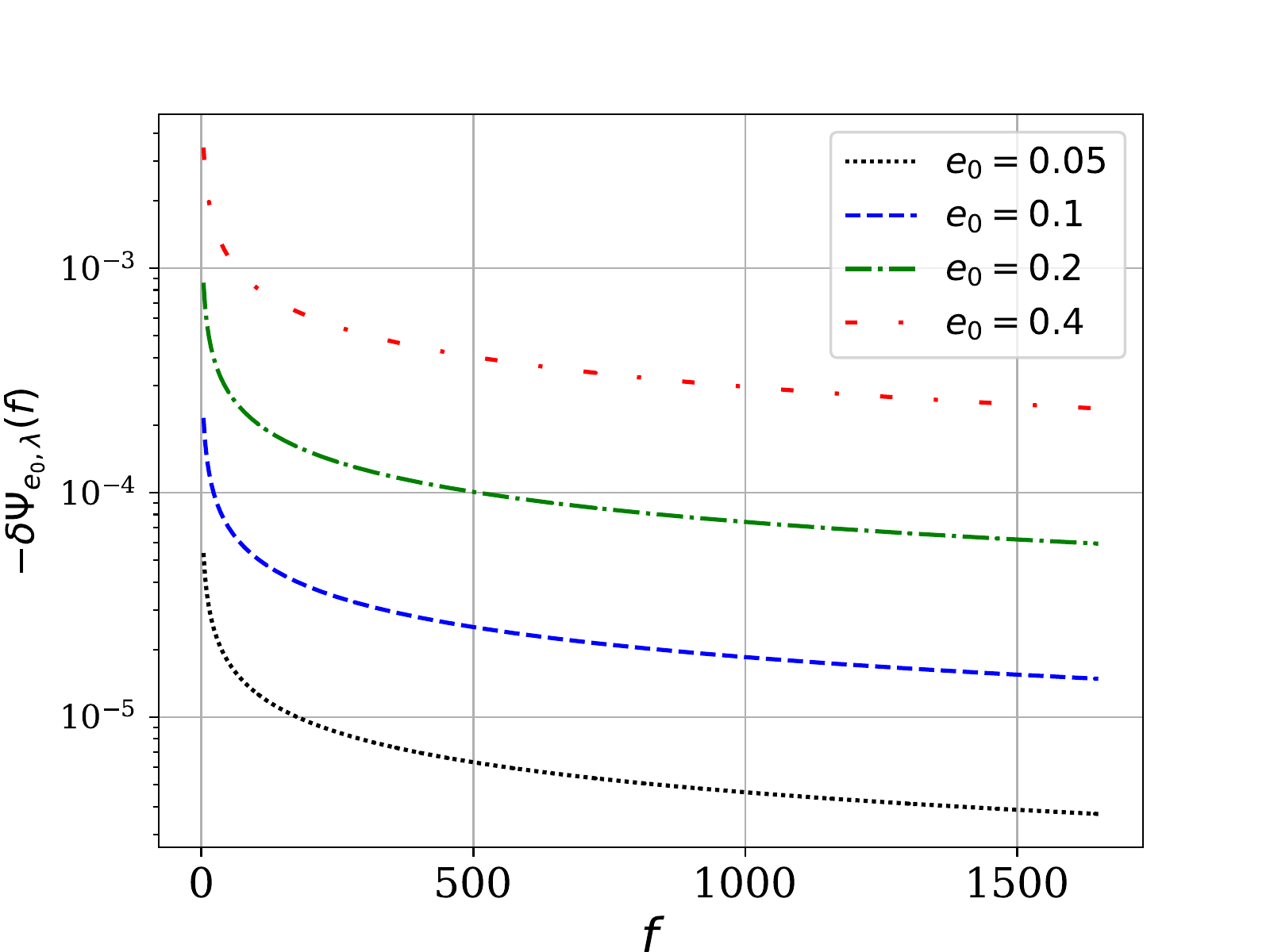}}
\subfigure[~$m_1=m_2=1.4 M_\odot$ and $q=1$ for $\Lambda=600$]{\label{}\includegraphics[width=85mm]{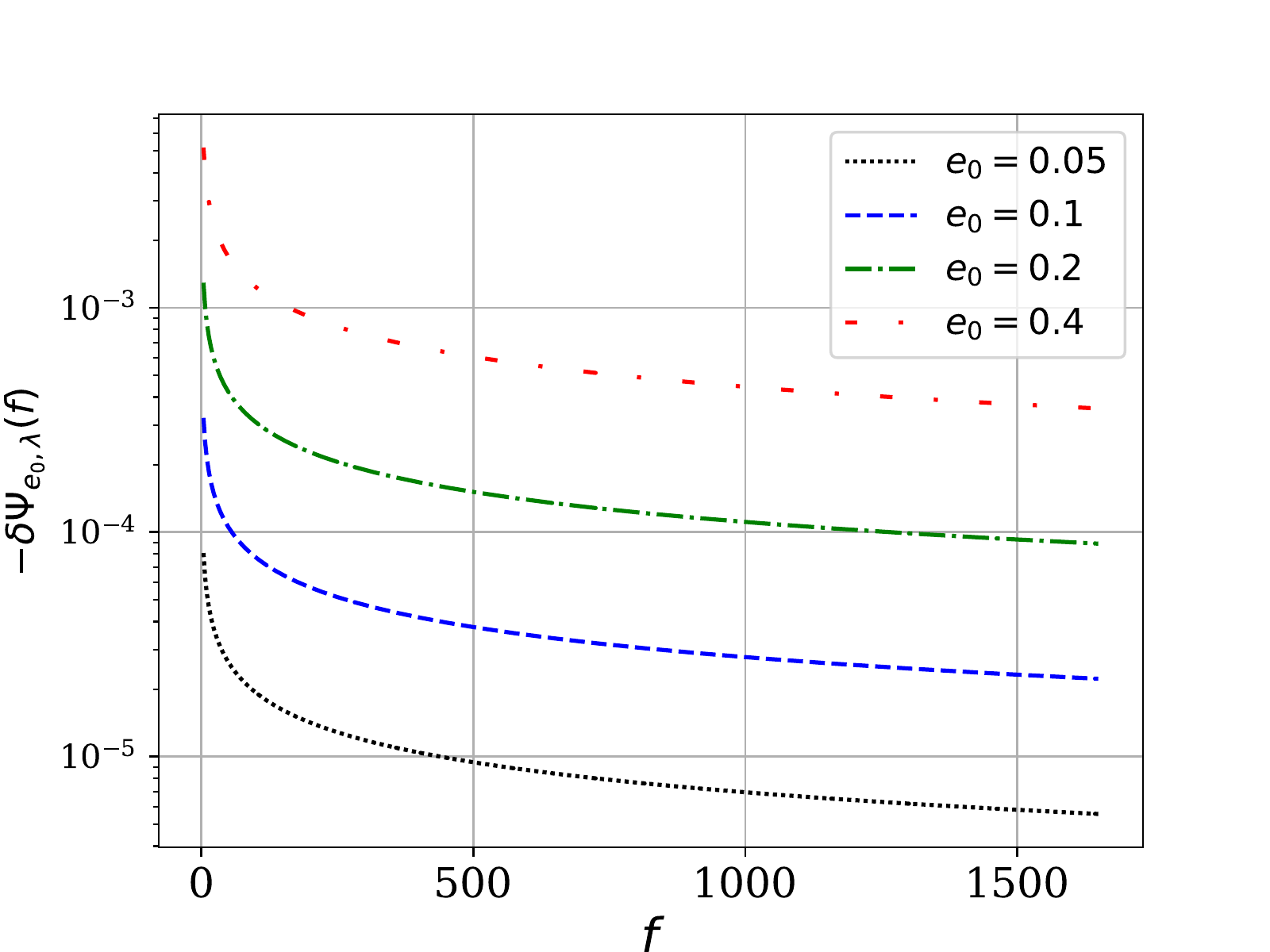}}

\caption{\small{We show the phasing $-\delta \Psi_{e_0,\lambda}$ (in radian) for a binary with two $1.4 M_\odot$ neutron stars as a function of gravitational wave frequency $f$ in the range $f_i =4 \text{Hz}$ to $f_{ISCO}$. The initial eccentricities taken for the figure are $e_0 = 0.05$, $0.1$,  $0.2$, and  $0.4$ with $\Lambda_1 = \Lambda_2 =\Lambda=400$ and $600$ respectively.}}
\label{plot1} 
\end{figure*}

To compute average over any quantity, say $X(\phi)$,  we average over orbital time period $(T)$. Then using the relation between time and $\phi$ this averaging can be translated over to an average over $\phi$ as follows \cite{PhysRev.131.435, Moore:2016qxz}:
\begin{equation}
    \langle X\rangle \equiv \int_0^T \frac{dt}{T}X=(1-e^2)^{3/2}\int^{2 \pi}_0 \frac{d \phi}{2 \pi} (1+e \cos \phi)^{-2} X(\phi).
\end{equation}
This, as a result, will give us the expression of the energy of the system at an inspiral instant \cite{Flanagan:2007ix}.

\begin{equation}\label{38}
    E=-\frac{1}{2} \mu M^{2/3}\omega^{2/3} \left(1- \gamma\left(3+\frac{55}{2} e^2\right)\right).
\end{equation}

For $e=0$, equation (\ref{38}) matches with Ref. \cite{Flanagan:2007ix} in static tide limit.

The energy emitted through GW can be computed from the time variation of the total quadrupole moment of the system,

\begin{equation}
    Q^T_{ij} = Q_{ij} +\mu x_i x_j.
\end{equation}

The total power radiated from the system can be obtained using the time derivatives of quadrupole moment in the following manner:
\begin{equation}\label{40}
    \Dot{E}=-\frac{1}{5}\left\langle\frac{d^3 Q^T_{ij}}{dt^3}\frac{d^3 Q^T_{ij}}{dt^3}-\frac{1}{3}\frac{d^3 Q^T_{ii}}{dt^3}\frac{d^3 Q^T_{jj}}{dt^3}\right\rangle,
\end{equation}
where $\langle\rangle$ represents the orbital average, and the repeated indices have been summed over.
Substituting the equations for $Q^T_{ij}$, we obtain:

\begin{equation}\label{45}
    \dot E=-\frac{32}{5} \mu^2 M^{4/3} \omega^{10/3}\left[1+2 \gamma \left(2+\frac{M}{m_2}\right) +e^2 \bar{P}_1 +e^4 \bar{P}_2 \right],
\end{equation}
where,
\begin{equation}\label{46}
\begin{split}
    \bar{P}_1 &=\frac{157}{24}+\gamma
    \left(\frac{371}{3}  +\frac{279}{4} \frac{M}{m_2}\right)\\
    \bar{P}_2 &=\frac{605}{32}+\gamma
    \left(\frac{436895}{672}  +\frac{11835}{16} \frac{M}{m_2}\right).
\end{split}
\end{equation}

With the emission of a gravitational wave, the binary loses its energy and inspirals inward. This, as a result, changes the energy of the system in the radiation reaction time scale. Along with modifying the energy, this also changes the eccentricity of the orbit. Therefore, to compute the impact of the inspiral, it is required to know the evolution of eccentricity to the GW frequency. Since the tidal interactions affect the GW flux, it also affects the eccentricity evolution. However, we can ignore it because we are focusing only on the leading order effect. The eccentricity can be expressed in terms of the frequency and initial eccentricity $e_0$. In the leading order it follows  \cite{Moore:2016qxz},

\begin{equation}\label{50}
    e(f) = e_0 \left(\frac{f_0}{f}\right)^{19/18},
\end{equation}
where $e_0$ is the eccentricity at $f=f_0$.

At this point, we are equipped to compute the phase contribution due to tidal effects. The expression of energy, flux, and eccentricity can be used to find the phase of the Fourier transform of the GW signal using the following equation:

\begin{equation}\label{54}
    \frac{d^2 \Psi}{d f^2}= \frac{2 \pi }{\dot E} \frac{dE}{df}
\end{equation}

This equation can be integrated using the expression of energy, flux, and the eccentricity evolution. As a result, it is possible to express the phasing in a series expansion of GW frequency $(f)$. We find the dephasing due to the tidal effect as follows:

\begin{widetext}
    
\begin{equation}\label{58}
\begin{split}
    \delta \Psi_T=\frac{-3}{128 \ \mu M^{2/3}\ f^{5/3}\ \pi^{5/3}}\Bigg[\Big(\frac{24 m_2 \pi^{10/3} f^{10/3} (11m_2+M)}{\mu M^{8/3}} + \frac{45 m_2 \pi^{10/3} f_0^{19/9} e_0^2 (378m_2+523 M)f^{11/9}}{26 \mu M^{8/3}}\Big)\lambda_1+1\leftrightarrow 2\Bigg],
\end{split}
\end{equation}
\end{widetext}
where $T$ in $\Psi_T(f)$ represents the tidal part of the phase. The first term is similar to the result found in Ref. \cite{Flanagan:2007ix} corresponding to the circular orbits. The second term represents the correction due to eccentricity. Defining the mass-ratio $q= \frac{m_1}{m_2}$, we can express the phase as:

\begin{widetext}
    
\begin{equation}
    \label{eq:phase in q}\delta \Psi_{e_0,\lambda}=\frac{-135 \pi^{5/3} f_0^{19/9} e_0^2 }{3328 (1+q)^2\ \mu^2 M^{4/3}\ f^{4/9}\ } \left[\left(901+523q\right)\lambda_1 + \left(901 q^2 + 523q\right)\lambda_2 \right].
\end{equation}

\end{widetext}

This expression will be used in the next sections to compute the impact on the GW.

\section{Observational impact}

In the previous section, we explored the orbital motion in the presence of a tidal field. From the derived expressions, we found the expression of the phase modification of the emitted GW of the system. In this section, we will focus on different types of systems, where these contributions can have relevance.

\subsection{Binary neutron star}

Binary neutron star (BNS) sources are most interesting in the context of tidal deformability. From the observation of GW170817 and GW190425, it has been possible to measure dimensionless tidal deformability $(\Lambda_i m_i^5 \equiv \lambda_i)$ and, as a result, put a constraint on $\Lambda_{1.4M_{\odot}}$ \cite{LIGOScientific:2018cki, Biswas:2021paf}. This, as a result, has shed light on the EOS of NS matter. It has also been possible to put an upper bound on the eccentricity of these systems \cite{Romero-Shaw:2020aaj, Lenon:2020oza}. From the theoretical understanding, as well as from the limited observations, we expect BNS systems to have very small eccentricities. It is expected that these systems will get circularized before entering the observable bands. Therefore, we do not expect the contribution of eccentricity-tide coupling in Eq. (\ref{eq:phase in q})
to contribute significantly. However, in a few recent studies, the possibility of forming eccentric compact binaries in the LIGO band by resonant and hierarchical triple and quadruple systems in globular clusters has been explored \cite{Wen:2002km, Seto:2013wwa, Antonini:2017ash, Rodriguez:2018jqu, Samsing:2017rat, Samsing:2017xmd, Liu:2018vzk, Hoang:2017fvh}. In Ref. \cite{East:2012xq}, dynamically captured binaries have also been explored. Although these events could be rare, they contain information about the formation channels, environment, and distribution.
With the advent of third-generation detectors such as the Einstein telescope and cosmic explorer, there is a possibility of measuring small values of dephasing at large signal-to-noise ratio (SNR). Therefore it might be possible to observe BNSs with small non-zero eccentricity, shedding light on the possible formation channels. For this purpose, we will study these systems here.

\begin{figure*}[ht]
\centering     %%% not \center

\subfigure[~$m_1=m_2=1.4 M_\odot$ and $q=1$ for $\Lambda=400$ ]{\label{}\includegraphics[width=85mm]{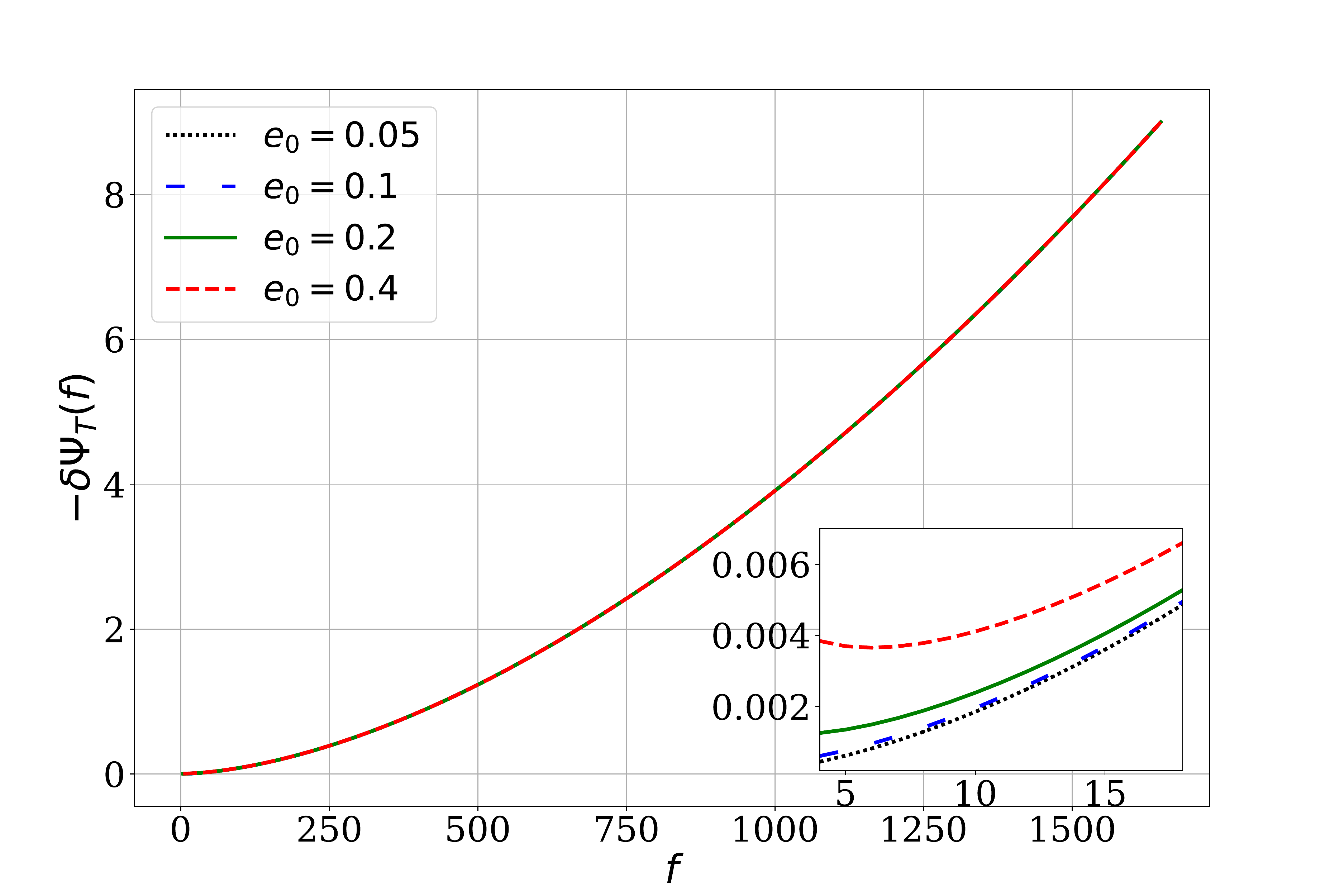}}
\subfigure[~$m_1=m_2=1.4 M_\odot$ and $q=1$ for $\Lambda=600$]{\label{}\includegraphics[width=85mm]{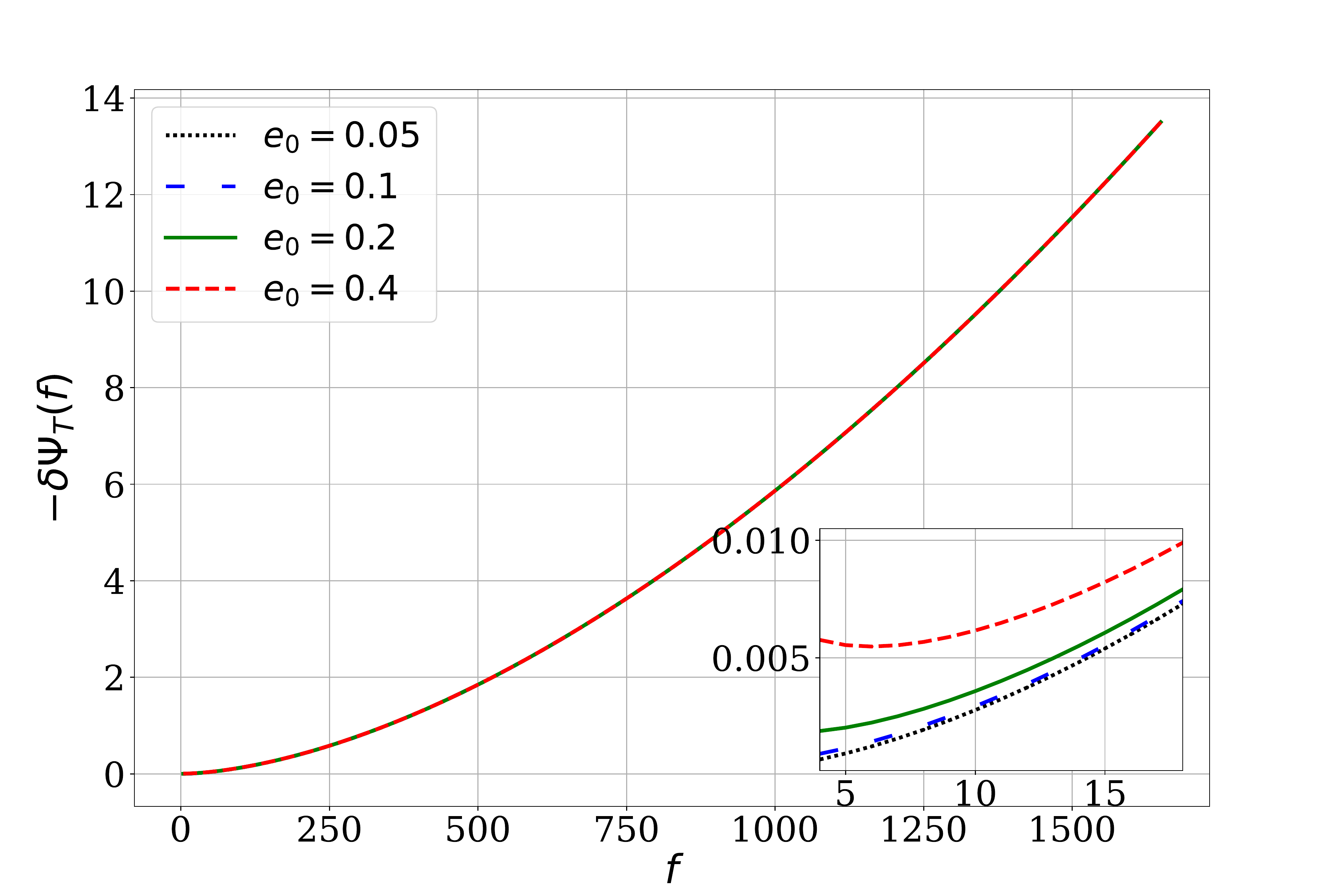}}

\caption{\small{We show the total tidal phasing $-\delta \Psi_{T}$ (in radian) for a binary with two $1.4 M_\odot$ neutron stars as a function of gravitational wave frequency $f$ in the range $f_i =4 \text{Hz}$ to $f_{ISCO}$. The initial eccentricities taken for the figure are $e_0 = 0.05$, $0.1$,  $0.2$, and  $0.4$ with $\Lambda_1 = \Lambda_2 =\Lambda=400$ and $600$ respectively.}}
\label{plot2} 
\end{figure*}

To study the impact, we plot $\delta \Psi_{e_0,\lambda}$ from Eq. (\ref{eq:phase in q}). The systems have been considered to be comprised of two equal mass NSs with $m_1=m_2=1.4 M_\odot$. The initial eccentricities $e_0$ of the systems are $e_0=0.05$, $0.1$, $0.2$, and $0.4$. Initial eccentricity has been defined at $f_0=10 \ \text{Hz}$. The frequency interval for the plots is from $f_i=4\text{Hz}$ up to $f_{ISCO}$, which refers to the frequency at the innermost stable circular orbit (ISCO). We used Ref.~\cite{Favata:2021vhw} 
 for the expression of ISCO frequency in BNS system. The figures have been represented in terms of dimensionless Love number $\lambda_i=\Lambda_i(m_i)^5$. The $\Lambda$ values taken for the system are $\Lambda=400$ and $600$.

The plots for the eccentricity-tide coupling term of the gravitational wave phase, $\delta \Psi_{e_0,\lambda}$, for $\Lambda=400$ as a function of gravitational wave frequency $f$ is shown in Fig. \ref{plot1} (a). From the plots, it can be seen that taking larger gravitational wave frequencies correspond to a smaller phasing. This is because the eccentricity of a binary is not a fixed quantity. It decreases with increasing frequency. This results in an overall decreasing pattern of the phase.  It can also be seen that larger $e_0$ corresponds to a larger phase difference, such as for $e_0=0.4$ and $0.2$. On the other hand, the contribution to the phase difference for small values of $e_0$, such as for values $0.05$ and $0.1$, is very small. This is also understandable since $\delta \Psi_{e_0,\lambda} \propto e_0^2$. In addition, not only does the coupling contribution decreases for larger gravitational wave frequencies, but also the value of phase is very small even at earlier frequencies. This implies that the $e_0$ has smaller contributions to the tidal perturbation as a whole.

\begin{figure}[ht]
\centering     %%% not \center  
\label{}\includegraphics[width=85mm]{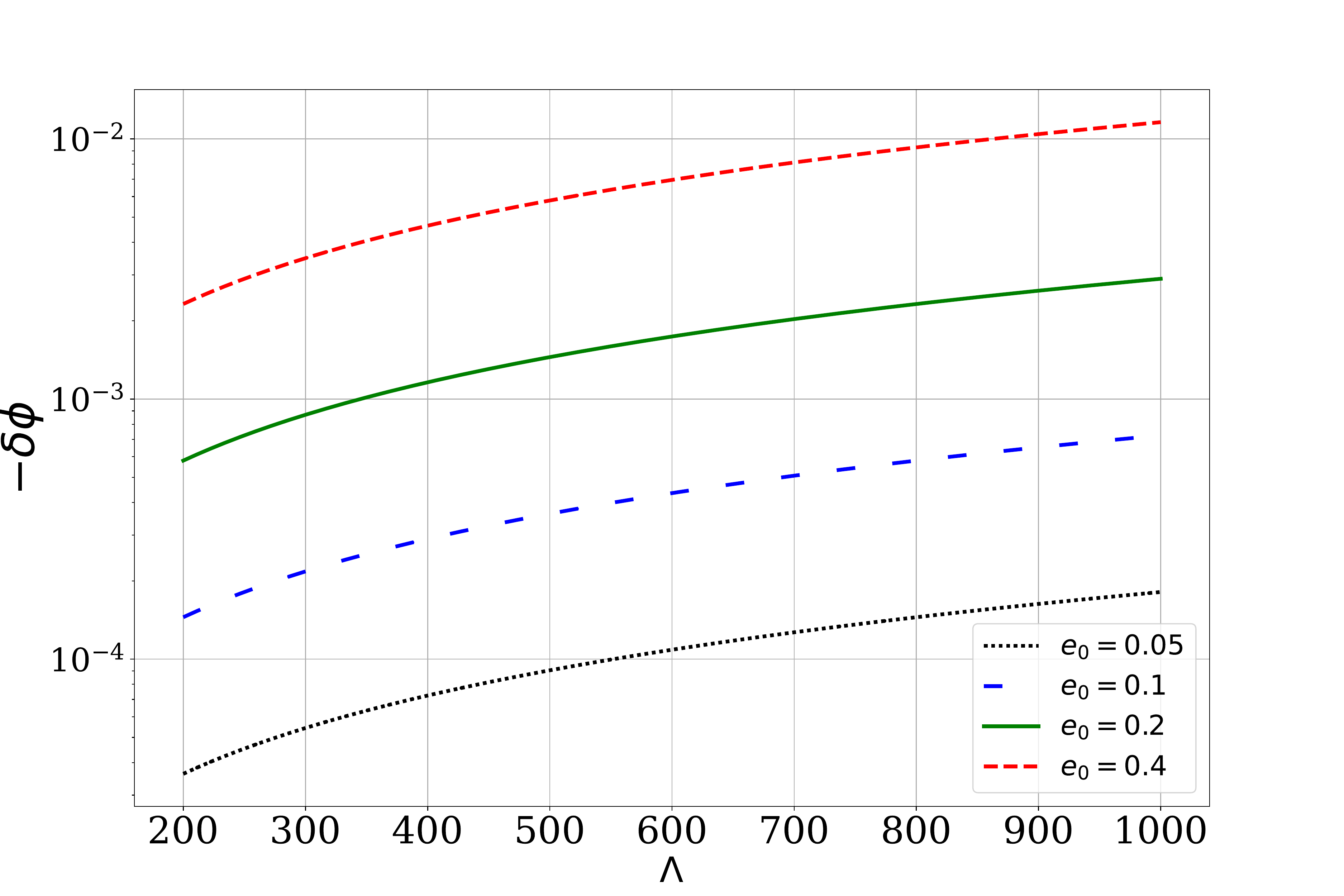}
\caption{\small{Shown above is the plot of accumulated dephasing $-\delta \phi$  in radian against dimensionless tidal deformability $\Lambda_1 = \Lambda_2 =\Lambda$  for  a binary of identical non-spinning neutron stars of $1.4 M_\odot$. The initial eccentricities taken for this system are $e_0 = 0.05$, $0.1$,  $0.2$, \text{and}  $0.4$ with $f_0=10 \text{Hz}$ for the frequency range $f_i=4 \text{Hz}$. }}
\label{plot3} 
\end{figure}

In Fig. \ref{plot1} (b), we show the plots for $\Lambda=600$. From the plots, we can see that similar observations apply, where the large values of initial eccentricities correspond to a larger contribution to the phase difference of the gravitational wave. However, when comparing to $\Lambda=400$, we can see that for $\Lambda=600$, the tidal perturbation contribution is higher for the same values of $e_0$. This is because $\delta \Psi_{e_0,\lambda} \propto \Lambda$, however, overall the contributions are very small.

In Fig. \ref{plot2},  we plot for $\delta \Psi_T$ to compare the effect of eccentricity with the whole phase due to tidal deformability. We plot for (a) $\Lambda=400$ and (b) $\Lambda=600$ and initial eccentricities $e_0=0.05$, $0.1$, $0.2$, and $0.4$. The values of $e_0$ appear to have smaller contributions to the total tidal phase, especially on the larger frequencies. Even larger values of $e_0$, which initially have a notable contribution, as shown in Fig. \ref{plot1} turn out to have smaller contributions comparatively. However, we see a notable increase on the slope from $\Lambda=400$ to $\Lambda=600$, which is consistent from Fig. \ref{plot1}, where a larger $\Lambda$ has a larger effect on the phase difference. It implies that $e_0$ only has significant contributions when taking smaller gravitational wave frequencies. Otherwise the overall structure has similar behavior as found in Ref. \cite{PhysRevD.77.021502}, which shows the tidal perturbations as a function of gravitational wave frequency for circular orbits.

\begin{figure}[ht]
%\centering     %%% not \center

\includegraphics[width=85mm]{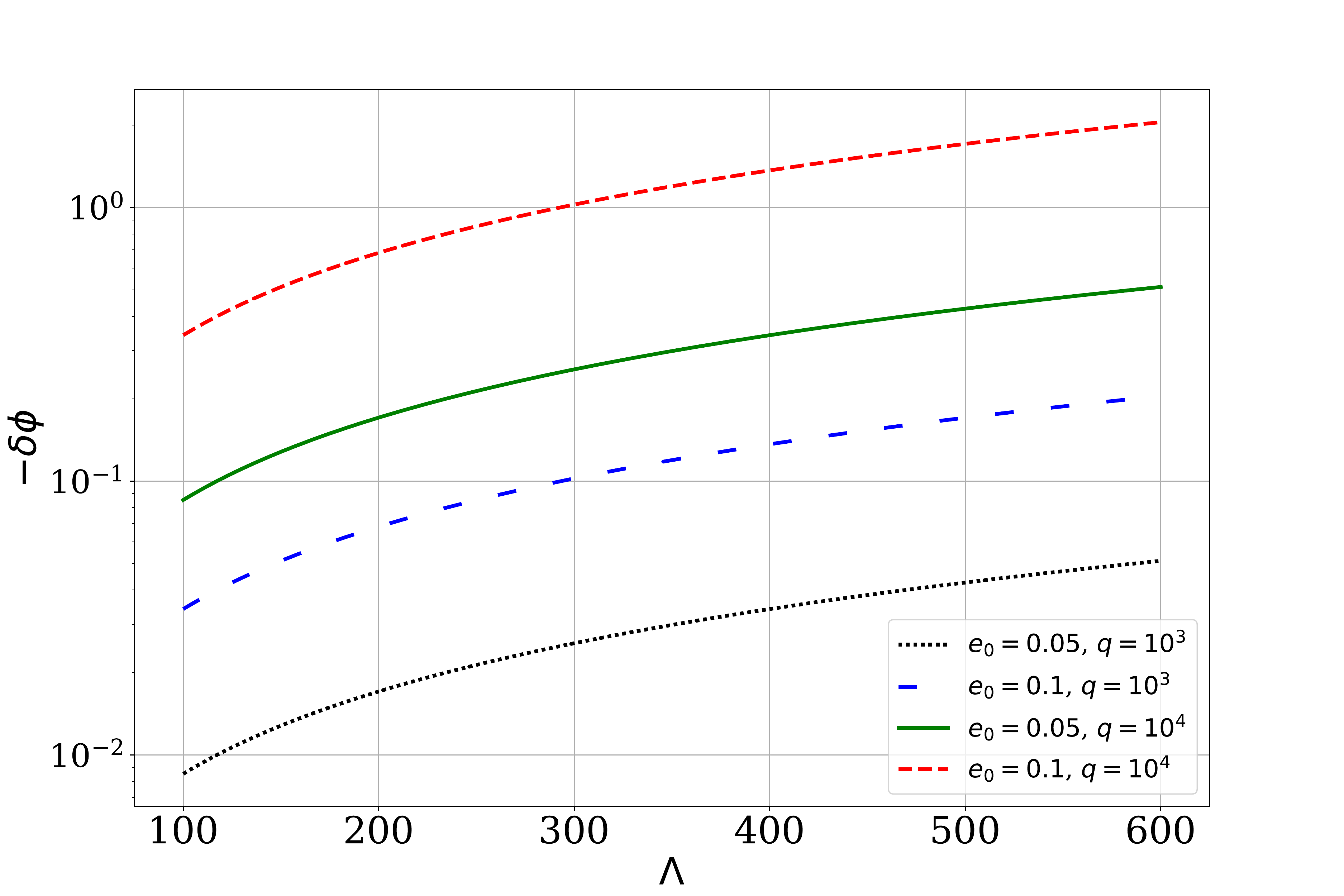}
\caption{\small{In the above figure, we show the accumulated dephasing $-\delta \phi$ (in radian) as a function of $\Lambda$ of a non-spinning NS in a neutron star-subsolar mass black hole binary. NS's mass is $M=1.4 M_\odot$ and the considered mass ratios are $q=10^3$ and $10^4$. The initial eccentricities are $e_0 = 0.05$ and $0.1$ with $f_0=10 \text{Hz}$. We computed the accumulation for the frequency range $f_i=4 \text{Hz}$ to $ f_{ISCO}$. }}
\label{plot4} 
\end{figure}

In Fig. \ref{plot3} we demonstrate the accumulated dephasing $-\delta \phi$ as a function of $\Lambda$ for an equal mass non-spinning binary neutron stars. This represents the total dephasing accumulated in a given frequency band. This is given by:
\begin{equation}\label{eq: accumulated phase}
    \delta \phi = \int^{f_{ISCO}}_{f_i} f \ df \frac{d^2 \delta \Psi_{e_0, \lambda}(f)}{df^2}.
\end{equation}

For this, we have considered $f_i = 4$ Hz. The mass of the individual stars is $m_1=m_2=1.4 M_\odot$ with a $\Lambda$ range of $200$ to $1000$. From the diagram, we can see that the accumulated dephasing increases with increasing $\Lambda$. This is accurate with our previous plots, where the dephasing is larger for $\Lambda=600$ than for $\Lambda=400$. In addition, the dephasing also increases for larger $e_0$, where $e_0=0.4$ is shown to have a larger contribution to the dephasing when compared to other values of $e_0$. These results precisely reflect the previous results for the dephasing of BNSs. However, the accumulated change is not very high. Dephasing contribution of an effect is indistinguishable from the absence of the effect in the context of scientific measurement if $\delta\phi^2 \leq 1/2\rho^2$, where $\rho$ is the SNR \cite{Flanagan:1997kp,Lindblom:2008cm}. Therefore, dephasing $\delta\phi$ is unobservable if the source does not cross a threshold SNR $\sim (\sqrt{2}\delta\phi)^{-1}$. From Fig. \ref{plot3}, this leads to a required SNR $\sim 7000$ to observe eccentricity-tide coupling for $e_0 =.05$. Therefore, this effect can be observed only with the systems with large eccentricity and very large SNRs in third-generation detectors or beyond.

\subsection{Neutron star-primordial black hole binary}

Primordial black holes (PBHs) could have formed in the early phase of the Universe due to the gravitational collapse of large overdensity regions \cite{Zeldovich:1967, Hawking:1971, Carr:1974, Chapline:1975}, e.g. arising from inflation \cite{Carr:1993aq, Ivanov:1994pa, Garcia-Bellido:1996mdl, Kim:1996hr}. It can possibly explain the fraction of the dark matter (DM). There exist astrophysical and cosmological constraints and clues on their abundance \cite{Carr:2016drx, Carr:2009jm, Carr:2020gox, Carr:2020xqk, Green:2020jor, Carr:2019kxo}. However, these limits and observations are highly model dependent. As a result, the status of PBHs to explain all the DM and GW observations is controversial. Unlike stellar black holes (stellar BHs), there is no physical process preventing the formation of PBHs lighter than the Chandrasekhar mass \cite{Chandrasekhar:1931ih} or in the pair-instability mass gap \cite{Rakavy:1967, Fraley:1968}. Therefore, if these objects do exist in nature, it is not unlikely that they can get captured in GW binaries. In this section, we will explore the impact of subsolar mass PBHs in a binary around an NS.

We have demonstrated in the last section that in a BNS, this effect will be hard to measure unless the system has very high eccentricity or/and SNR. However, if the system has a large mass-ratio, the larger body's contribution dominates, and it is $\propto q$ in the leading order. For this purpose, we explore the impact of this effect on binaries comprising a $1.4 M_\odot$ NS and sub-solar mass black holes (smBHs).

From Eq. (\ref{eq: accumulated phase}), we generate plots for the accumulated dephasing of a non-spinning neutron star-subsolar mass black hole binary with the frequency range $f_i=4 \text{Hz}$ to $f_{ISCO}$ \cite{Taracchini:2013wfa}. We consider both of the bodies to be non-spinning. The results are shown in Fig. \ref{plot4}. The initial eccentricities taken for these plots are $e_0=0.05$ and $0.1$ with a $\Lambda$ range of $100$ to $600$. From both plots, it can be seen that a larger $e_0$ corresponds to a larger dephasing effect, which is consistent with the results from other systems. In addition, when comparing both plots, we can see that a larger $q$ also corresponds to a larger dephasing effect, which is expected. We find that even for $e_0 \sim .05$ accumulated dephasing can be $\sim \mathcal{O}(.1)$ radian, requiring an SNR $\gtrsim 7$ to be observable.

Therefore, if an extreme mass-ratio inspiral of this kind is observed then eccentricity-tide coupling will be a measurable effect. However, the capture rates of these systems are likely to be very low \cite{Capela:2013yf, Genolini:2020ejw}. During the capture of an smBH, it is likely that the smBH will disrupt the structure of the NS and accrete matter from it. In such a case, observing these binaries may not be possible.

\begin{figure*}[ht]
\centering     %%% not \center

\subfigure[~$M=60 M_\odot, \ q=10^4$]{\label{}\includegraphics[width=85mm]{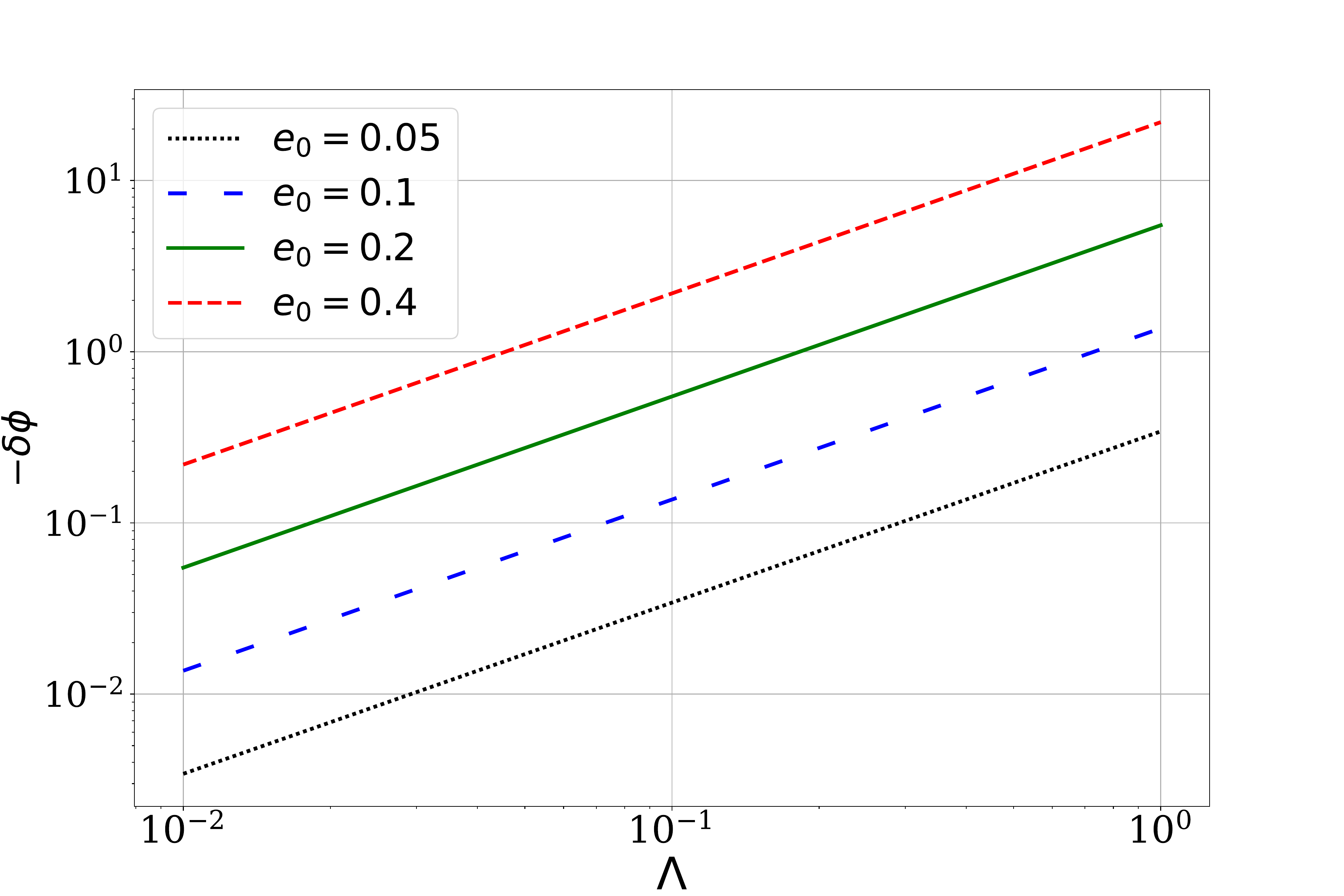}}
\hfill
\subfigure[~$M=100 M_\odot, \ q=10^4$]{\label{}\includegraphics[width=85mm]{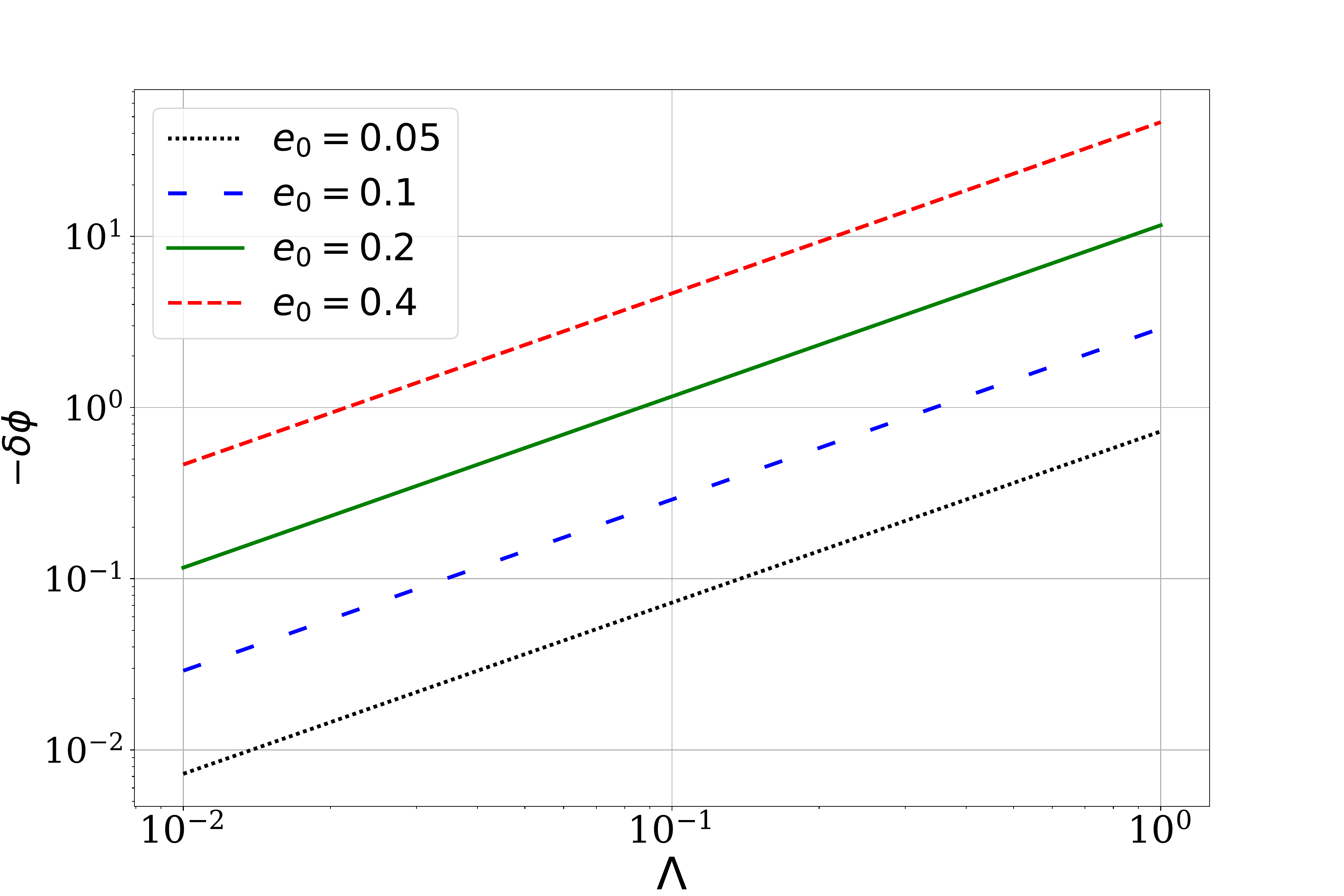}}

\caption{\small{The plots above shows  the accumulated dephasing $-\delta \phi$ (in radian) as a function of $\Lambda_1 = \Lambda$ in the range $10^{-2}$ to $1$ for non-spinning stellar black hole-subsolar mass black hole binary. The initial eccentricities taken for these systems are $e_0 = 0.05$, $0.1$,  $0.2$, and  $0.4$ with $f_0=10 \text{Hz}$. The accumulation is computed from $f_i=4 \text{Hz}$ to $f_{ISCO}$.}}
\label{plot5} 
\end{figure*}

\subsection{Stellar mass black hole-primordial black hole binary}

In general relativity (GR), the emergence of new physics is expected to appear in the strong-field regime, when approaching the UV cutoff. Planck-scale modifications of black hole horizons and modification of BH structure have been proposed in several works as resolutions to the information-loss paradox. Among the models of exotic compact objects (ECOs), there are BH polymers, fuzzballs, Firewalls, gravastars, Boson stars etc. to mention a few \cite{Cardoso:2016rao}. BHs in GR have a vanishing Love number \cite{Binnington:2009bb, Landry:2014jka,LeTiec:2020bos, Chia:2020yla}. However, deviation from the classical BH picture in GR introduces non-zero Love numbers. Therefore, the measurement of $\Lambda \neq 0$ will be a smoking gun for deviation from this description.

In the previous sections, we have demonstrated that the eccentricity-tide coupling is a small effect. Since this effect is $\propto \Lambda$, it contributes even lesser for smaller values. However the effect can become stronger if the systems have a large mass-ratio, as discussed in the previous section. Here, we demonstrate the effect of eccentricity-tide coupling for small values of $\Lambda$ when the binary comprises a non-spinning stellar mass and a subsolar mass BH.

Fig. \ref{plot5} shows the dephasing for a non-spinning stellar black hole-subsolar mass black hole binary accumulated in the frequency range $f_i = 4$ Hz and $f_{ISCO}$ \cite{Taracchini:2013wfa} as a function of $\Lambda$ in the range $10^{-2}$ to $1$. The total masses for these binaries are (a) $M=60 M_\odot$ and (b) $M=100 M_\odot$ respectively with $q$ value fixed at $10^4$. The initial eccentricities taken for this system are $e_0 = 0.05$, $0.1$,  $0.2$, and  $0.4$ with an initial frequency  $f_0=10$ Hz. In these systems the effect can have reasonable dephasing $(\mathcal{O}(.1))$ radian even for $\Lambda\sim .1$ and $e_0 \sim .1$. For $e_0 \sim .4$ it can even be $\sim \mathcal{O}(10)$ radian.

\subsection{Extreme mass-ratio inspiral}

In this section, we explore the extreme mass-ratio inspiral (EMRIs) comprising of spinning supermassive bodies (SMBs). Extreme mass-ratio inspirals are binaries with very different masses of bodies. We explored this kind of systems in the last sections. However, in this section, we focus on EMRIs with SMBs and stellar mass bodies. The frequency of emitted GW by these systems falls in the milliHertz band. These sources will be observed with the future space-based Laser Interferometer Space Antenna (LISA) \cite{LISA:2017pwj}. Due to the mass-ratio, it will be possible to measure small values of $\Lambda$ of the SMBs \cite{Pani:2019cyc, Datta:2021hvm, Piovano:2022ojl}. It will lead to very accurate and precise testing of the nature of the SMBs. Furthermore, these systems are expected to have large eccentricities. Therefore, if the bodies deviate from the classical BH paradigm, the impact of eccentricity-tide contribution can be significant.

\begin{figure*}[ht]
\centering     

\subfigure[~$M=10^6 M_\odot, q=10^5$ ]{\label{}\includegraphics[width=85mm]{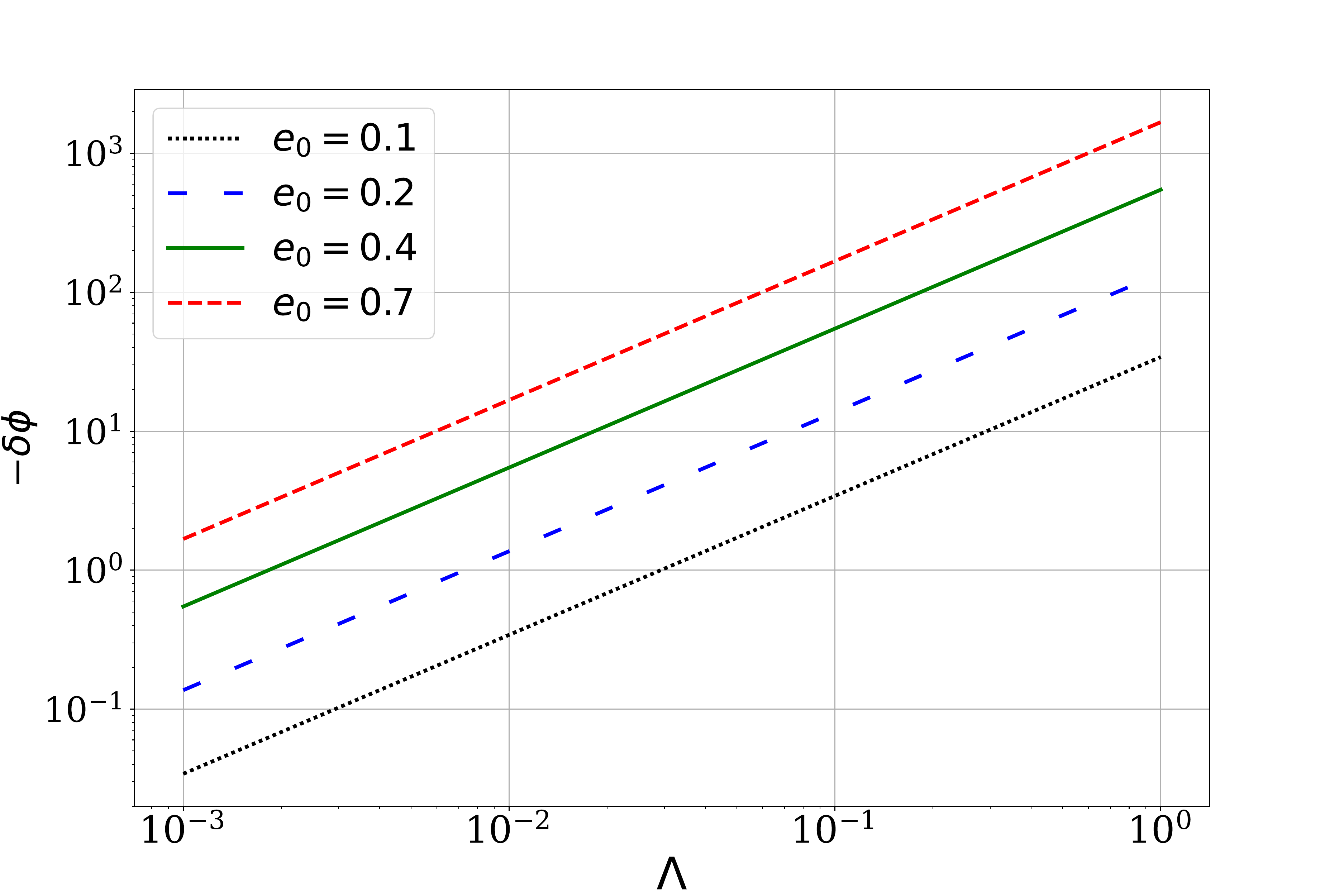}}
\subfigure[~$M=10^6 M_\odot, q=5 \times 10^5$ ]{\label{}\includegraphics[width=85mm]{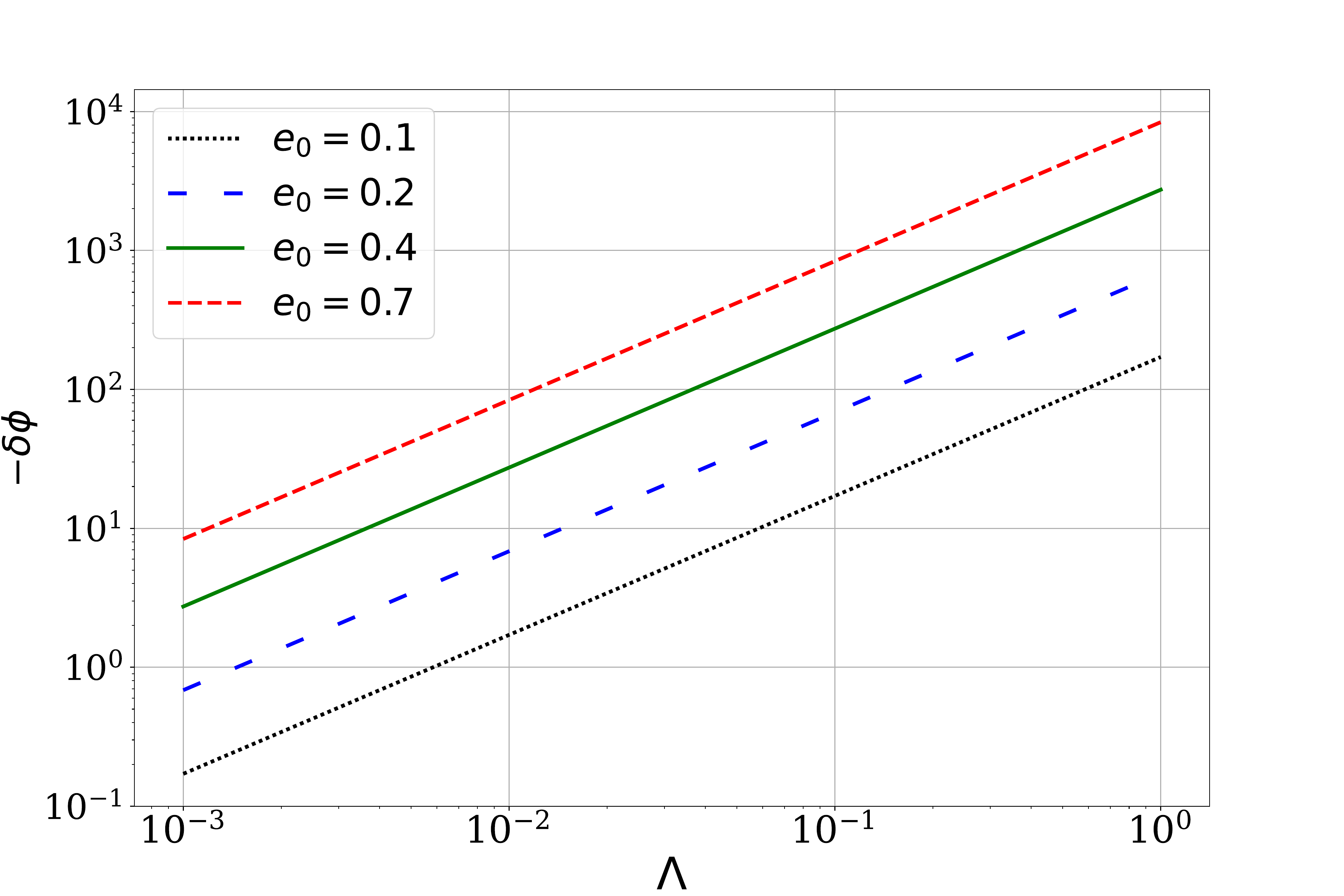}}

\caption{\small{The figure above shows the accumulated dephasing $-\delta \phi$ (in radian)  against $\Lambda_1 = \Lambda$ for extreme mass-ratio inspirals with $M=10^6 M_\odot$. The initial eccentricities we use are $e_0 = 0.1$, $0.2$,  $0.4$, \text{and}  $0.7$ with $f_0=1\text{mHz}$. The SMB has spin $\chi=0.8$. The accumulation is computed for the frequency range $f_i=.4 \text{mHz}$ to $f_{ISCO}$.}}
\label{fig: EMRI_SMB} 
\end{figure*}

To analyze the contribution of eccentricity to the phase difference of EMRIs, we use Eq. (\ref{eq: accumulated phase}) and generate plot for $-\delta \phi$ as a function of $\Lambda$ accumulated in the frequency range $f_i=  .4\text{mHz}$ to $f_{ISCO}$ \cite{Taracchini:2013wfa}. They are deomnstrated in Fig.\ref{fig: EMRI_SMB}. We assumed  SMB's spin to be $0.8$ which has been used to compute the $f_{ISCO}$. The initial eccentricities $e_0$, defined at $f_0=1$mHz, used for these systems are $e_0 = 0.1$, $0.2$,  $0.4$, \text{and}  $0.7$. The total mass of the extreme mass-ratio inspirals are fixed at $M=10^6 M_\odot$ with mass-ratio (a) $q=10^5$ and (b) $5 \times 10^5$, representing a $10M_{\odot}$ BH and a $2M_{\odot}$ NS, respectively. As expected, the dephasing can be large even for $\Lambda\sim 10^{-3}$. Note, although we assumed $e_0^2$ to be small to begin with, we have demonstrated the impact of large eccentricity $e_0=0.7$. This is just to show the largeness of the eccentricity-tide coupling in these systems. We find that the the accumulated dephasing can be $\sim\mathcal{O}(10)$ radian even for $\Lambda\sim 10^{-3}$ and reasonable $e_0$. However, to properly model the effect for high eccentricity systems, much more rigor will be needed in the future.

\section{discussion and conclusion}

In this paper, we studied the impact of the tidal deformability of a star in an eccentric binary. Assuming a static tide, we found the orbital equations of motion. We assumed the orbits to be eccentric in the absence of the tidal field. Then, we assumed that in the presence of the tide, the orbital motion will change perturbatively. Under such an assumption, we found the solution of the equation of motion in the leading order in eccentricity for the first time. With this newly found result, we computed the total energy of the system and energy flux emitted as GW. Then, we finally calculated the expression of the phase. We found that at the leading order, the phase is proportional to $e_0^2 \Lambda$. Due to the nature of the expression, we call it eccentricity-tide coupling. To our knowledge, this is the first time it has been demonstrated that there will be direct coupling between initial eccentricity $e_0$ and the tidal deformability $\Lambda$. In the limit $e_0=0$, we recover the result for the tidal contributions in circular binaries, found in \cite{Flanagan:2007ix}.

With the result at hand, we plotted the phase for BNSs. We find that the contribution is small for realistic values of eccentricities. We also computed the accumulated dephasing of the system and concluded that it will be measured only for large eccentricities and very large SNRs in third-generation detectors. We also computed the accumulated dephasing in EMRIs. Due to the large mass-ratio, we found that the impact of the effect in the accumulated dephasing becomes larger. We observe that the effect will be measurable if a binary comprising NS and a primordial BH is observed.

Classical BHs in GR have vanishing Love numbers. Measuring a nonzero Love number, therefore, indicates a deviation from the general relativistic BH paradigm. Hence, we computed the impact of eccentricity-tide coupling, assuming small values of tidal deformability for supposed BHs. EMRIs with stellar mass BH and smBH have a reasonable contribution to dephasing for realistic eccentricities. Finally, we focus on EMRIs with SMBs at the center. These systems will be observed with the future space-based detector LISA. In these systems, the impact is significant even for very small $\Lambda$ of SMBs. Since EMRIs are expected to have large eccentricities, large inclinations, and high precession, it will be important to find the contribution of tidal effects in generic orbits targeting these kinds of systems. This, as a result, will provide us a plethora of information regarding the nature of the supermassive compact objects.

Since we have considered the static tide, we ignored the impact of the mode excitations explored in Ref. \cite{Xu:2017hqo, Yang:2019kmf, Wang:2020iqj}. It will be interesting to model such effects analytically so that it becomes useful in the observational context. We leave such an endeavor for the future.

\section*{Acknowledgement}

We thank K.G. Arun and Huan Yang for reading the manuscript and providing us with valuable inputs. We thank Bhaskar Biswas, Rahul Dhurkunde, Shilpa Kastha, Alexander H. Nitz, Khun Sang Phukon, and  Paolo Pani for helpful discussions. 

%We thank the AEI, Hann
%

%\newpage
%\bibliographystyle{unsrt}
\bibliography{References.bib}

\end{document}